\documentclass[aps,prl,twocolumn,showpacs,amsmath,amssymb,superscriptaddress]
{revtex4-1}
\usepackage{graphicx}
\usepackage{color}
\begin{document}
\title{Classification of Phase Transitions by 
Microcanonical Inflection-Point Analysis}
\author{Kai Qi}
\email{k.qi@fz-juelich.de}
\affiliation{Theoretical Soft Matter and Biophysics, Institute of Complex 
Systems and Institute for Advanced Simulation, Forschungszentrum J\"ulich, 
52425 J\"ulich, Germany}
\affiliation{Soft Matter Systems Research Group, Center for Simulational
Physics, Department of Physics and Astronomy, The University of Georgia, 
Athens, GA 30602, USA}
\author{Michael Bachmann}
\email{bachmann@smsyslab.org; http://www.smsyslab.org}
\affiliation{Soft Matter Systems Research Group, Center for Simulational
Physics, Department of Physics and Astronomy, The University of Georgia, 
Athens, GA 30602, USA}
\begin{abstract}
By means of the principle of minimal sensitivity
we generalize the microcanonical inflection-point analysis method by 
probing derivatives of the microcanonical entropy
for signals of transitions in complex systems.
A strategy of systematically identifying and locating
independent and dependent phase transitions of any order is proposed.
The power of the generalized method is demonstrated in applications 
to the ferromagnetic Ising model and a coarse-grained model for
polymer adsorption onto a substrate. The results shed new 
light on the intrinsic phase structure of systems with cooperative behavior.
\end{abstract}
\maketitle 
Conventionally, the identification of phase transitions is based 
on locating non-analyticities, discontinuities, or divergences in 
thermodynamic variables (e.g., entropy, pressure, magnetization) and response 
functions (specific heat, compressibility, susceptibility, etc.), 
respectively. These quantities can be represented by distinct 
derivatives of appropriate thermodynamic potentials. According to 
Ehrenfest's classification scheme~\cite{ehrenfest1}, the order 
of the transition is determined by the lowest derivative which 
exhibits catastrophic behavior at the transition temperature. 
However, this can only occur in the thermodynamic limit. 
Thermodynamic quantities describing the macrostate of 
finite systems do not show any such obvious transition behavior. 
Therefore, the rapidly growing interest in understanding
thermodynamic activity in finite systems, such as nanoscale 
systems relevant for biology and modern nanotechnology, necessitate a 
generalized identification and classification scheme for phase transitions. 
Microcanonical 
statistical 
analysis~\cite{Thirring1970,Janke1998,Kastner2000,Gross2001,dunkel1,%
Junghans2006} has turned out to be a useful 
basis 
for first systematic schemes~\cite{Schnabel2011,BachmannBook2014}.

In this context, it has been common to analyze first-order-like transitions 
in finite systems 
by means of Maxwell's 
construction, where the backbending region in the transition 
regime of the energetic temperature curve is replaced by a 
flat segment. However, Maxwell construction only applies to 
single transitions of first order and can neither be used if the transition is 
composed or accompanied by subphase transitions~\cite{kb1}, nor if it is of 
higher order. 
However, by replacing the ``flatness'' idea of Maxwell's 
construction by the more general principle of minimal 
sensitivity~\cite{Stevenson1981}, these issues can be resolved as will be 
discussed in the following.

The principle of minimal sensitivity (PMS) was proposed 
to solve the ambiguity of results obtained by applying different 
renormalization schemes (RS) in conventional perturbation 
theory~\cite{Stevenson1981,Stevenson1981-2}. It asserts that if a 
truncated perturbation 
expansion in some RS 
depends on unphysical parameters, of which the exact result must be 
independent, the parameter values should be chosen so as 
to minimize the sensitivity of the approximant to small variations 
in those parameters. The PMS has found numerous 
applications~\cite{Kleinert2009,Kauffmann1984,Duke1982,Akeyo1993,Canet2005,
Wrigley1983,Stevenson1983,Inui2006,Buckley1993,Lu2007,Canet2003}.

In this Letter, we show that the combination 
of microcanonical inflection-point analysis~\cite{Schnabel2011} and the PMS 
enable the systematic
identification, characterization, and classification of first- and higher-order 
transitions in complex systems of any size. The 
analysis reveals surprising transition features, 
suggesting the discrimination of regular (\emph{independent}) transitions from 
\emph{dependent} transitions, which only exist in combination with a regular 
transition. Unexpectedly, even the  
two-dimensional ferromagnetic Ising model exhibits signals of transitions 
other than the established single second-order phase transition. 
Furthermore, the difficulty in uniquely identifying the compact phases in the 
long-standing problem of polymer adsorption can be traced back to a complex 
structure of subphases separated by higher-order transitions. By employing the 
generalized analysis method 
proposed here, we obtain novel results for these systems.

\begin{figure}
\centerline{\includegraphics[width=8.8cm]{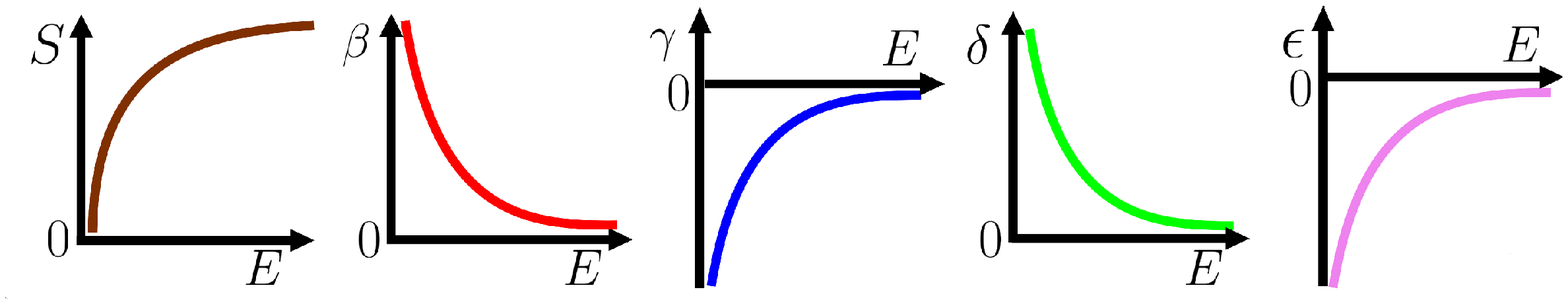}}
\caption{\label{Fig1}%
Typical monotony of microcanonical entropy $S(E)$ and 
first- to fourth-order derivatives $\beta=dS/dE,\ldots,\epsilon=d^4S/dE^4$, if 
no transition occurs.}
\end{figure}

The microcanonical entropy, defined by $S(E)=k_{\mathrm{B}} \, \mathrm{ln} \, 
g(E)$, where $g(E)$ is the density of states with system energy $E$, contains 
the complete 
information about the phase behavior of a system. In the thermodynamically  
relevant energetic region, it is 
a monotonically increasing concave function [Fig.~\ref{Fig1}]. Changes 
of the phase behavior are signaled by alterations of the curvature of $S(E)$ 
leading to characteristic monotonic features of the inverse microcanonical 
temperature, which is given by
\begin{equation}
\beta(E) \equiv \frac{d S(E)}{dE}. \label{beta}
\end{equation}
In energetic regions without transition signals, 
$\beta(E)$ is a strictly monotonically decreasing convex function
[Fig.~\ref{Fig1}]. 

Canonically, large fluctuations at 
the transition temperature lead to a dramatically increased expectation 
value of the system energy 
$\langle E \rangle$, signaling a phase or pseudophase transition. In 
consequence, in microcanonical 
analysis, it is expected 
to occur in the energetic region, where the inverse temperature $\beta(E)$ 
responds 
\emph{least sensitively} to energy changes. However, in situations where 
transitions are not identified in $\beta(E)$, higher-order derivatives 
of the entropy might still reveal signals of cooperative behavior. Generally, 
if no 
transition 
occurs, the derivatives of $S(E)$ are either monotonically increasing 
concave or monotonically decreasing convex functions [Fig.~\ref{Fig1}] 
in the energetic regime where thermodynamic phase transitions can occur. 
A change in monotonicity causes an inflection 
point which we call 
\emph{inflection point of least sensitivity} if the derivative changes least 
upon variation in energy and provides a transition signal at this 
energy. 

\begin{figure}
\centerline{\includegraphics[width=8.8cm]{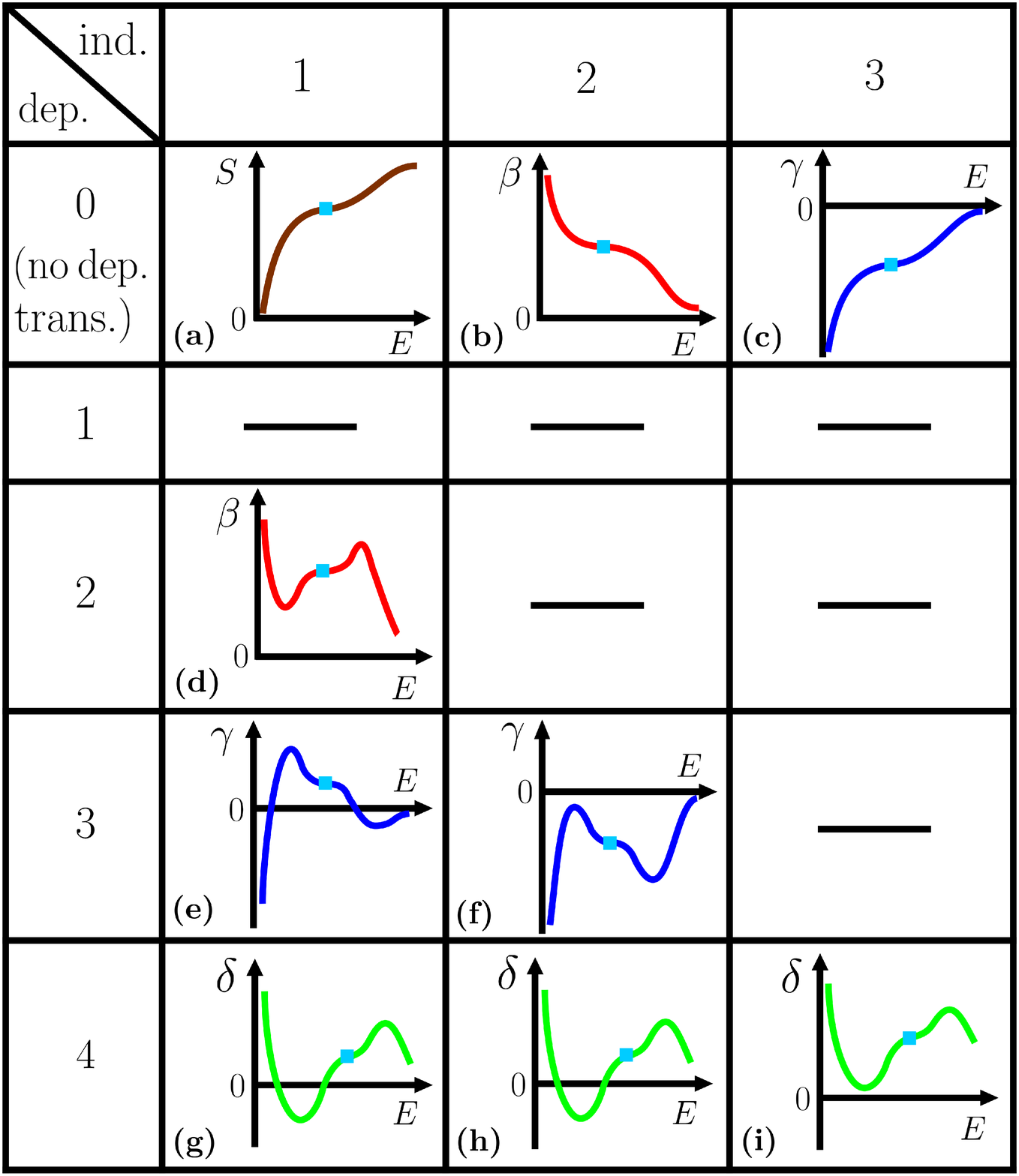}}
\caption{\label{Fig2}%
Entropy and lowest-order derivatives for independent and potential 
dependent 
transitions. Least-sensitive inflection points are marked, but are not 
associated with each other. If it occurs at all, a dependent transition is 
indicated by a least-sensitive inflection point at a higher energy than the 
independent transition it accompanies.}
\end{figure}

The typical first-order transition scenario is 
sketched in Fig.~\ref{Fig2}(a). For finite systems, the entropy possesses a 
convex region caused by surface effects~\cite{Gross2001,BachmannBook2014}. The 
slope of the unique double-tangent across the convex regime is the 
Gibbs-Maxwell hull and the
energy difference between the touching points of double-tangent and $S(E)$ 
defines the latent heat. In the thermodynamic limit, the convex 
``intruder'' disappears as surface effects become irrelevant and the slope of 
the Gibbs-Maxwell line corresponds to the inverse transition temperature. 
The convex region in $S(E)$ causes a ``backbending'' of the $\beta(E)$ 
curve. If we define the transition energy 
$E_{\mathrm{tr}}$ associated  
with the 
least sensitive inflection point in $S(E)$, $\beta(E)$ has a 
positive-valued minimum at $E_{\mathrm{tr}}$,
\begin{equation}
\beta(E_{\mathrm{tr}})>0,
\label{first}
\end{equation}
and $\beta(E_{\mathrm{tr}})$ is the inverse transition temperature.
Since the backbending region is formed directly in the otherwise monotonically 
decreasing convex $\beta(E)$ curve, the occurrence of the 
first-order transition 
is independent of the possible existence of other transition signals.

Consequently, we identify an independent second-order 
phase transition by an inflection 
point, where $\beta(E)$ is least sensitive to changes in energy [see 
Fig.~\ref{Fig2}(b)]. 
The corresponding derivative $\gamma(E)$ 
exhibits a negative-valued peak at the transition energy, i.e.,  
\begin{equation}
\gamma(E_{\mathrm{tr}})=\left. \frac{d^2S(E)}{dE^2} 
\right|_{E=E_{\mathrm{tr}}}<0. 
\label{second}
\end{equation}
Note that only signatures of first- and second-order transitions are 
directly visible in $\beta(E)$. However, if such transition 
signals are not found, the PMS condition can be applied to 
higher derivatives of $S(E)$ as well. For example, since $\gamma(E)$ is 
strictly concave if no transition occurs, 
an inflection point at which the $\gamma(E)$ curve behaves least 
sensitively signals an independent third-order phase 
transition [Fig.~\ref{Fig2}(c)]. The derivative of 
$\gamma(E)$ near 
$E=E_{\mathrm{tr}}$ 
forms a valley with a positive-valued minimum, i.e.,
\begin{equation}
\delta(E_{\mathrm{tr}})=\left. \frac{d^3S(E)}{dE^3} 
\right|_{E=E_{\mathrm{tr}}}>0. \label{delta}
\end{equation}

To generalize, we define an \emph{independent transition of odd order
$(2k-1)$} (where $k$ is a positive integer),
if there is a least-sensitive inflection point in the $(2k-2)$th derivative 
of $S(E)$ and the corresponding minimum in the $(2k-1)$th 
derivative of $S(E)$ is positive, i.e.,
\begin{equation}
\left. \frac{d^{(2k-1)}S(E)}{dE^{(2k-1)}} 
\right|_{E=E_{\mathrm{tr}}}>0. \label{odd_order_trans}
\end{equation}
Analogously, an \emph{independent transition of even order $2k$} 
($k$ is a positive integer) is associated with a least-sensitive inflection 
point in the $(2k-1)$th derivative of 
$S(E)$ and the corresponding negative-valued maximum in the $(2k)$th 
derivative 
of 
$S(E)$, i.e.,
\begin{equation}
\left. \frac{d^{2k}S(E)}{dE^{2k}} \right|_{E=E_{\mathrm{tr}}}<0. 
\label{even_order_trans}
\end{equation}
It is noteworthy that generalized inflection-point 
analysis also reveals another type of transitions. These 
\emph{dependent} transitions can only occur in coexistence with 
an independent transition of lower order (see Fig.~\ref{Fig2}).

According to our proposed classification scheme, a \emph{dependent 
transition of even order} $2l$ (where $l$ is a 
positive integer) exists, if there is a least-sensitive inflection point in 
the 
$(2l-1)$th derivative of $S(E)$ which can be identified by a positive-valued 
minimum in the $(2l)$th derivative in the transition region of the 
corresponding 
independent transition,
\begin{equation}    
\left. \frac{d^{2l}S(E)}{dE^{2l}} \right|_{E=E_\mathrm{tr}^\mathrm{dep}}>0. 
\label{even_order_dep_trans}
\end{equation}
Consequently, a \emph{dependent transition of odd order} ($2l+1$) (with $l$ 
being positive integer) is indicated by 
a least-sensitive inflection point in the $2l$th derivative of $S(E)$ and is 
characterized by a negative-valued maximum in the $(2l+1)$th derivative:
\begin{equation}
\left. \frac{d^{(2l+1)}S(E)}{dE^{(2l+1)}} 
\right|_{E=E_\mathrm{tr}^\mathrm{dep}}<0. \label{odd_order_dep_trans}
\end{equation}
It is important to note that the occurrence of an independent 
transition \emph{does not necessarily} imply the existence of dependent 
transitions, whereas the opposite is true. Least-sensitive inflection points 
indicating dependent transitions are not simply a consequence of the 
monotonic shape associated with the curve of a derivative of $S(E)$ that 
features an independent transition. As Fig.~\ref{Fig2} shows, for example, a 
first-order independent transition can be (but does not necessarily have to 
be) accompanied by a dependent transition of any order higher than~1.

We leave it to future work to determine 
the circumstances for dependent transitions to exist and their scaling 
properties in 
the thermodynamic limit. Since dependent transitions always occur at a 
higher energy than the corresponding independent transition, the 
former can be interpreted as a precursor of the latter in the less ordered 
phase. This might be of interest in applications in materials science as 
the dependent transitions indicate instabilities in an otherwise stable 
phase. This relationship sheds 
new light on our general understanding of ordering principles leading to 
phase transitions.

We now demonstrate the power of this novel method.
In the first example, we re-analyze the ferromagnetic-paramagnetic phase 
transition of the two-dimensional 
Ising model on a square lattice.
The energy of a spin configuration is given by $E=-J\sum_{\langle ij\rangle} 
s_i s_j$, where
$s_i=\pm 1$ represents the possible spin orientations. Only 
nearest-neighbor spin pairs contribute. The energy scale, given 
by the coupling 
constant $J$, is set to unity. The extraordinary advantage of this model 
is that it has been solved rigorously~\cite{Onsager1944,Kaufman1949}. The 
microcanonical entropy $S(E)$ can be obtained exactly for any system 
size~\cite{Beale1996} and allows for a direct application and test of our 
method.

\begin{figure}
\centerline{\includegraphics[width=8.0cm]{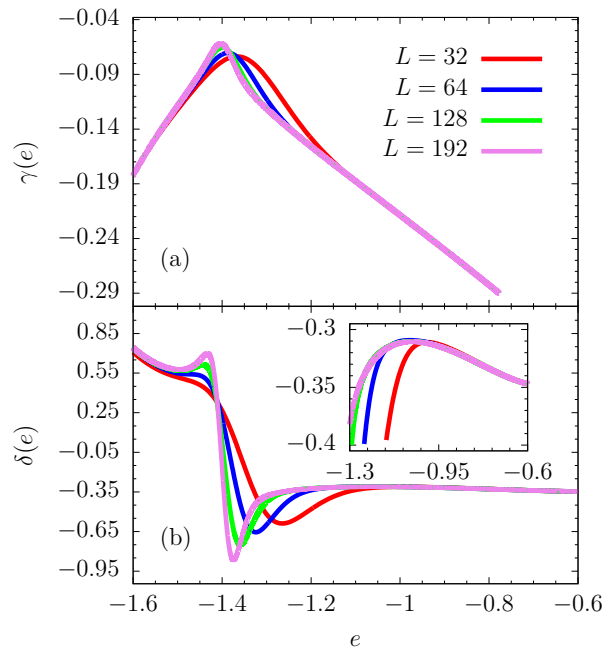}}
\caption{\label{Fig3}%
Derivatives of microcanonical entropies (a) $\gamma(e)$; (b) $\delta(e)$ 
of the ferromagnetic Ising model on a square lattice for various 
system sizes $L=32,64,128,$ and $192$ as functions of $e=E/L^2$.}
\end{figure}

To compare the results for different system sizes $L$, 
we introduce energies and entropies per spin $e=E/L^2$ and 
$s=S/L^2$, respectively. Figure~\ref{Fig3} shows the $\gamma$ and $\delta$
derivatives of $s(e)$ and, as expected, strong second-order transition signals
are indicated by the negative-valued maxima in $\gamma(E)$ for all system 
sizes studied. The transition signal becomes 
more pronounced 
with increasing system size. Remarkably, the value of $\delta(e)$ is 
independent of system size and one is reminded of the Binder 
cumulant crossings~\cite{binder1}. It is obvious that a microcanonical 
scaling analysis 
is worth being tested (note that the peak value of $\gamma(e)$ must converge 
to zero 
in the thermodynamic limit), but this is left to future work.

More interesting in the given context is the revelation of additional 
transition signals shadowing the well-known independent second-order phase 
transition 
at $e_\mathrm{tr}\approx -1.403$ ($T\approx 2.276$). At the lower energy 
$e_\mathrm{tr}\approx 
-1.492$ ($T\approx 2.235$), $\delta(e)$ exhibits a positive-valued minimum for 
$L=128$ and $192$, 
which corresponds to an independent transition of third order (for the smaller 
systems $L=32$ and $64$, it is of fourth order). Furthermore, the inset in 
Fig.~\ref{Fig3}(b) reveals a negative peak at higher energies 
$e_\mathrm{tr}^\mathrm{dep}\approx -1.057$ ($T\approx 2.561$) for all 
systems studied. It features 
an additional dependent third-order transition in the paramagnetic phase. 
\begin{figure}
\centerline{\includegraphics[width=8.8cm]{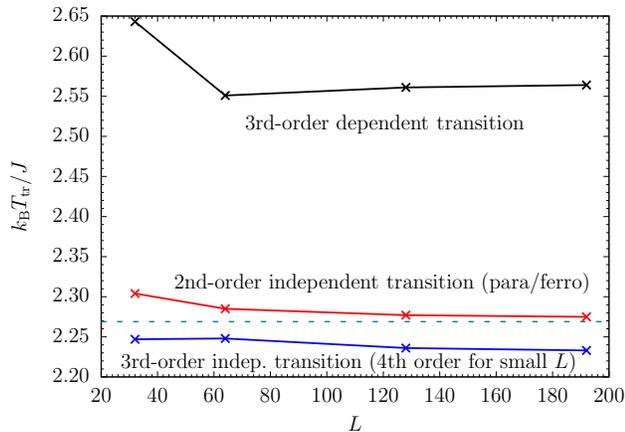}}
\caption{\label{newFig4}%
Transition temperatures of  
the 2D Ising model identified by our method for various systems sizes (lines 
are guides to the eye). The dashed line marks the Onsager solution for the 
critical point in the thermodynamic limit, 
$T_\mathrm{tr}=2J/k_\mathrm{B}\ln(1+\sqrt{2})$.}
\end{figure}

It is worth noting that all transition signals 
become more significant with increasing system size, implying that 
the two higher-order transitions may also exist 
in the thermodynamic limit. As Fig.~\ref{newFig4} shows, the 
transition temperatures remain well separated for larger systems. A thorough 
scaling analysis is needed and more 
detailed 
studies are necessary to characterize their nature, though. Due to their lower 
significance compared to the critical transition, it is likely that in 
all previous studies their effects have been absorbed in corrections-to-scaling 
of the power laws of the critical transition and, hence, remained undetected.

\begin{figure}
\centerline{\includegraphics[width=8.8cm]{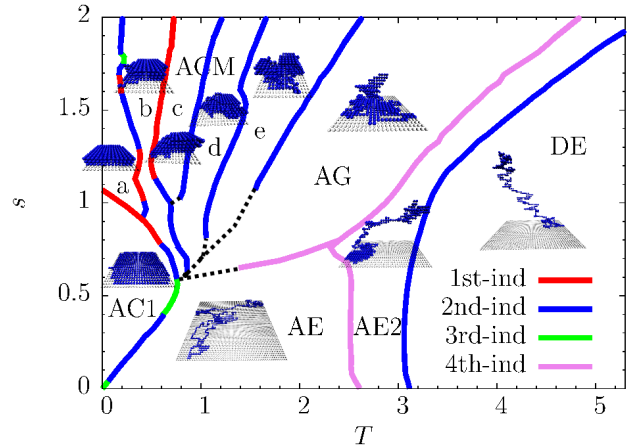}}
\caption{\label{newFig5}%
Phase diagram of a grafted polymer interacting 
with an adhesive substrate (only lines of independent transitions are shown). 
Representative conformations 
in the respective phase regions are also shown. Dotted lines 
correspond to transition lines of higher than fourth order.}
\end{figure}

As a second example, we study a coarse-grained model of a grafted lattice 
polymer interacting with an 
adhesive surface. The energy of the 
system can be written as $E(n_s,n_m)=-n_s-s n_m$, where 
$n_s$ and $n_m$ denote the
numbers of nearest-neighbor monomer-substrate contacts and
nearest-neighbor non-bonded monomer-monomer contacts, 
respectively~\cite{Bachmann2006}. The dimensionless reciprocal
solubility $s$ effectively controls the quality of the implicit solvent.  
Simulations were performed using the contact-density 
chain-growth algorithm, which yields the 
number of states for given $(n_s,n_m)$ pairs~\cite{BachmannBook2014}. This 
so-called contact density can be transformed into the density of states for any 
given value of $s$ without additional simulations. By means of generalized 
inflection-point analysis, transition signals are located and classified, 
and the microcanonical transition temperatures identified. Accumulating this 
information, we can construct the $T$-$s$ hyperphase diagram. For a polymer 
with 503 monomers, it is shown in Fig.~\ref{newFig5}. At high 
temperatures, 
the polymer is desorbed and expanded (DE) in the free space. Below the 
second-order adsorption transition, larger sections of 
the polymer get adsorbed onto the substrate (AE2). 

Under sufficiently good solvent conditions, the 
polymer forms fully adsorbed 
and expanded conformations (AE), 
whereas adsorbed globular (AG) or crystalline and 
multi-layered (ACM) structures dominate otherwise. 
In the ACM subphases, the compactness of the polymer 
is divided into different levels. 
The ACMa subphase contains the most compact conformations, whereas the 
compactness of the structures is the least in the ACMe subphase.
In the conformational phase AC1, 
the polymer forms a compact, filmlike single layer on the substrate. 
The identification of the compact phases has been a long-standing problem, but 
our novel analysis method provides a unique approach with no room for 
ambiguities. It should be noted that the simulation of larger systems in the 
compact phases is extremely difficult, but conventional scaling analyses of the 
adsorption-desorption transition yield promising results~\cite{pmb1}.  

In the novel statistical analysis method introduced in this Letter, least 
sensitive inflection points in the microcanonical entropy and its 
derivatives are used as indicators of phase and pseudophase transitions. 
The hierarchical classification scheme applies to two different classes of 
transitions, which we call independent and dependent. Dependent 
transitions can only exist in combination with a lower-order independent 
transition and may be interpreted as precursors of the latter. As a proof of 
concept for the power of the method we studied the 
two-dimensional 
Ising model, which revealed additional higher-order transitions in the vicinity 
of the critical transition, and polymer adsorption. In the latter example, the 
complete hyperphase diagram in solubility-temperature space could be 
constructed, 
which helps understand better the structure of the compact phases. An in-depth 
discussion of the intriguing details is future work.

The methodology presented here is versatile and promising as it can be 
universally applied to complex physical systems of any size. The consequently 
hierarchical scheme significantly advances previous methods in 
identifying and classifying phase and pseudophase transitions and is 
particularly useful for applications in the emerging field of 
complex systems on mesoscopic scales with high cooperativity, for which no 
thermodynamic limit exists.

This work has been supported by the NSF under Grant No.\ DMR-1463241.


\begin{thebibliography}{99}
%
\bibitem{ehrenfest1}
P.~Ehrenfest, \emph{Phasenumwandlungen im ueblichen und erweiterten Sinn, 
classifiziert nach den entsprechenden Singularitaeten des thermodynamischen 
Potentiales}, in: Proc.\ Royal Acad.\ Amsterdam (Netherlands), 
Vol.~\textbf{36} (1933), p.~153; Comm.~Leiden Suppl.\ No.~75b.
%
\bibitem{Thirring1970}
W.~Thirring, Z.~Phys.\ \textbf{235}, 339 (1970).
%
\bibitem{Janke1998}
W.~Janke, Nucl.\ Phys.~B \textbf{63A-C}, 631 (1998).
%
\bibitem{Kastner2000}
M.~Kastner, M.~Promberger, and A.~H\"uller, J.~Stat.\ Phys.\ \textbf{99}, 
1251 (2000).
%
\bibitem{Gross2001}
D.~H.~E.\ Gross, \emph{Microcanonical Thermodynamics} (World Scientific, 
Singapore, 2001).
%
\bibitem{dunkel1}
J.~Dunkel and S.~Hilbert, Physica A \textbf{370}, 390 (2006).
%
\bibitem{Junghans2006}
C.~Junghans, M.~Bachmann, and W.~Janke, Phys.\ Rev.\ Lett.\ \textbf{97}, 
218103 (2006).
%
\bibitem{Schnabel2011}
S.~Schnabel, D.~T.\ Seaton, D.~P.\ Landau, and M.~Bachmann, Phys.\ 
Rev.~E \textbf{84}, 011127 (2011).
%
\bibitem{BachmannBook2014} 
M.~Bachmann, \emph{Thermodynamics and Statistical Mechanics of 
Macromolecular Systems} (Cambridge University Press, Cambridge, 2014).
%
\bibitem{kb1}
T.~Koci and M.~Bachmann, Phys.\ Rev.~E \textbf{95}, 032502 (2017).
%
\bibitem{Stevenson1981}
P.~M.\ Stevenson, Phys.\ Rev.~D \textbf{23}, 2916 (1981).
%
\bibitem{Stevenson1981-2}
P.~M.\ Stevenson, Phys.\ Lett.~B \textbf{100}, 61 (1981).
%
\bibitem{Kleinert2009}
H.~Kleinert, \emph{Path Integrals in Quantum Mechanics, Statistics, 
Polymer Physics, and Financial Markets}, 5th ed.\ (World Scientific, Singapore, 
2009).
%
\bibitem{Kauffmann1984}
S.~K.\ Kauffmann and S.~M.\ Perez, J.~Phys.~A: Math.\ Gen.\ \textbf{17}, 
2027 (1984).
%
\bibitem{Duke1982}
D.~W.\ Duke and J.~D.\ Kimel, Phys.\ Rev.~D \textbf{25}, 2960 (1982).
%
\bibitem{Akeyo1993}
J.~O.\ Akeyo and H.~F.\ Jones, Phys.\ Rev.~D \textbf{47}, 1668 (1993).
%
\bibitem{Canet2005}
L.~Canet, Phys.\ Rev.~B \textbf{71}, 012418 (2005).
%
\bibitem{Wrigley1983}	
J.~C.\ Wrigley, Phys.\ Rev.~D \textbf{27}, 1965 (1983).
%
\bibitem{Stevenson1983}
P.~M.\ Stevenson, Phys.\ Rev.~D \textbf{27}, 1968 (1983).
%
\bibitem{Inui2006}
M.~Inui, A.~Ni\'egawa, and H.~Ozaki, Prog.\ Theor.\ Phys.\ \textbf{115}, 
411 (2006).
%
\bibitem{Buckley1993}
I.~R.~C.\ Buckley, A.~Duncan, and H.~F.\ Jones, Phys.\ Rev.~D \textbf{47}, 
2554 (1993).
%
\bibitem{Lu2007}
W.~F.\ Lu, C.~K.\ Kim, and K.~Nahm, J.~Phys.~A: Math.\ Theor.\ 
\textbf{40}, 14457 (2007).
%
\bibitem{Canet2003}
L.~Canet, B.~Delamotte, D.~Mouhanna, and J.~Vidal, Phys.\ Rev.~D \textbf{67}, 
065004 
(2003).
%
\bibitem{Onsager1944}
L.~Onsager, Phys.\ Rev.\ \textbf{65}, 117 (1944).
%
\bibitem{Kaufman1949}
B.~Kaufman, Phys.\ Rev.\ \textbf{76}, 1232 (1949).
%
\bibitem{Beale1996}
P.~D.\ Beale, Phys.\ Rev.\ Lett.\ \textbf{76}, 78 (1996).
%
\bibitem{binder1}
K.~Binder, Z.~Phys.~B \textbf{43}, 119 (1981).
%
\bibitem{Bachmann2006}
M.~Bachmann and W.~Janke, Phys.\ Rev.~E \textbf{73}, 041802 (2006).
%
\bibitem{pmb1}
J.~A.\ Plascak, P.~H.~L.\ Martins, and M.~Bachmann, Phys.\ Rev.~E 
\textbf{95}, 050501(R) (2017).
%
\end{thebibliography}
\end{document}